\documentclass[conference]{IEEEtran}

\usepackage{graphicx}      
\usepackage{amsmath} 
\usepackage{amssymb}  
\usepackage{epsfig}
\usepackage{graphicx}
\usepackage[hang]{caption2}
\usepackage[normalsize,it,hang]{subfigure}
\usepackage[latin1]{inputenc}
\usepackage[greek,english]{babel}
\usepackage{longtable}
\usepackage{booktabs}

\newtheorem{remark}{Remark}[section]

\ifCLASSINFOpdf
\else
\fi

\begin{document}
\title{Preliminary remarks on \\ option pricing and dynamic hedging}

\author{\IEEEauthorblockN{Michel Fliess}
\IEEEauthorblockA{LIX (CNRS, UMR 7161),\\
\'Ecole polytechnique, 91128 Palaiseau, France \\
 {\tt \small Michel.Fliess@polytechnique.edu}}
\and
\IEEEauthorblockN{C\'edric Join}
\IEEEauthorblockA{INRIA -- Non-A \& CRAN (CNRS, UMR 7039),
\\Universit\'e de Lorraine, BP 239, 54506 Vand{\oe}uvre-l\`es-Nancy, France \\
{\tt \small cedric.join@univ-lorraine.fr}}}

\maketitle

\begin{abstract}
~ An elementary arbitrage principle and the existence of trends
in financial time series, which is based on a theorem published in 1995 by
P. Cartier and Y. Perrin, lead to a new understanding of option
pricing and dynamic hedging. Intricate problems related to violent
behaviors of the underlying, like the existence of jumps, become
then quite straightforward by incorporating them into the trends.
Several convincing computer experiments are reported.
\\ ~ \\
\hspace{0.1cm}
~~\textit{Keywords}--- Quantitative finance, option pricing, European option,
dynamic hedging, replication, arbitrage, time series, trends, volatility, abrupt changes,
model-free control, nonstandard analysis.
\end{abstract}

\IEEEpeerreviewmaketitle

\section{Introduction}
Option pricing intends like many other financial techniques to tame
as much as possible market risks. The Black-Scholes-Merton (BSM)
approach (\cite{bs,merton}), which is forty years old, is still by
far the most popular setting, although some of its drawbacks and
pitfalls were known shortly after its publication. It had an
enormous impact\footnote{The \emph{performative} aspect of the BSM
approach might also be stressed (see, \textit{e.g.},
\cite{performatif}).} on the huge development of modern quantitative
finance. Its heavy use of advanced mathematical tools, like
stochastic differential equations and partial differential
equations, explains to a large part the features of today's
mathematical finance, which is enjoying a great popularity not only
among academics but also among practitioners. Many textbooks (see,
\textit{e.g.},
\cite{chauveau,dana,karoui,franke,hull,lamberton,portait,wilmott}) provide an
excellent overview of this lively and fascinating field.

Let us add in the context of this conference that a growing number
of references exploits the connections of the BSM setting with
methods stemming from various engineering fields. We mention here:
\begin{itemize}
\item learning techniques (see, \textit{e.g.}, \cite{neural,lo}),
\item control theory
(see, \textit{e.g.},
\cite{barron,bernhard,cal,davis,mc,pham,primbs,vollert}).
\end{itemize}

In 1997, Scholes and Merton won the Nobel Prize in economics --
Black died in 1995 -- not for the discovery of the pricing formulas
which were already known (\cite{boness,samuelson,samuelmerton,sprenkle}), but for
the methods they introduced for deriving them.\footnote{See,
\textit{e.g.}, the historical comments in \cite{smith},
\cite{mehrling}, \cite{haug} and \cite{ht}.} The most elegant
concepts of \emph{replication} and \emph{dynamic delta hedging},
which are now central both in theory and practice, have nevertheless
been the subject of severe criticisms for their lack of realism
(see, \textit{e.g.}, \cite{derman}). Dynamic delta hedging moreover
cannot be extended to more general stochastic processes exhibiting
jumps for instance (\cite{merton1}).

Pricing formulas are derived here via an elementary arbitrage
principle which employs the expected return of the underlying and goes
back at least to \cite{bachelier} and \cite{bronzin}.\footnote{See
the comments by \cite{schacher2} and \cite{zimmer}.} Combined with
the utilization of \emph{trends} (\cite{fes}) it  permits to
\begin{enumerate}
\item alleviate one of the most annoying paradoxes in modern approaches
that concerns the uselessness of the expected return of the underlying
(see, \textit{e.g.}, \cite{black2}),
\item deal quite simply with more subtle behaviors of the underlying, which may
exhibit jumps, by incorporating those behaviors in the trends,
\item define a new more realistic dynamic hedging.
\end{enumerate}

Our paper is organized as follows. Section \ref{cp} summarizes and
sometimes improves some facts already presented earlier (see
\cite{agadir} and the references therein). Section \ref{arbi}
recalls how pricing formulas may be derived via an elementary
arbitrage principle, \textit{i.e.}, without replication. Section
\ref{wt} slightly modifies those formulas by taking trends into
account. A new dynamic hedging, which employs both the pricing
formulas and the trend of the underlying, is proposed in Section
\ref{dhed}. Due to an obvious lack of space, the convincing computer
illustrations, which are displayed in Section \ref{illu}, are
limited to a quite violent behavior of the underlying. Section
\ref{conclu} further analyzes the change of paradigm
which might arise from this new setting.


\section{The Cartier-Perrin theorem and some of its consequences: A short
review}\label{cp}
\subsection{Trend}
The theorem due to Cartier and Perrin \cite{cartier} is expressed in the language of
\emph{nonstandard analysis}. It depends on a time sampling
$\mathfrak{TS}$ where the difference $t_{\nu + 1} - t_\nu$ is
\emph{infinitesimal}, \textit{i.e.}, ``very small''. Then, under a
mild integrability condition, the price $S (t)$ of the financial
quantity may be decomposed (\cite{fes}) in the following way
\begin{equation*}\label{decomposition}
\boxed{S(t) = S_{\tiny{{\mathfrak{TS}}, {\rm trend}}} (t) +
S_{\tiny{{\mathfrak{TS}}, {\rm fluct}}} (t)}
\end{equation*}
where
\begin{itemize}
\item $S_{\tiny{{\mathfrak{TS}}, {\rm trend}}}$ is the \emph{trend}, or the \emph{mean}, or the \emph{average}, of $S$;
\item $S_{\tiny{{\mathfrak{TS}}, {\rm fluct}}}$ is a \emph{quickly fluctuating}
function around $0$, \textit{i.e.}, $\int_{\tau_0}^{\tau_1}
S_{\tiny{{\mathfrak{TS}}, {\rm fluct}}}(\tau) d\tau$ is
infinitesimal for any finite interval $[\tau_0, \tau_1]$,
\item $S_{\tiny{{\mathfrak{TS}}, {\rm trend}}}$ and
$S_{\tiny{{\mathfrak{TS}}, {\rm fluct}}}$ are unique up to an
additive infinitesimal quantity.
\end{itemize}
\begin{remark}$S_{\tiny{{\mathfrak{TS}}, {\rm trend}}} (t)$, which is ``smoother''
than $S(t)$, provides a mathematical justification (\cite{fes}) of
the \emph{trends} in \emph{technical analysis} (see, \textit{e.g.},
\cite{bechu,kirk}).
\end{remark}
\begin{remark}\label{trendagadir}
Note that $S_{\tiny{{\mathfrak{TS}}, {\rm fluct}}} (t)$ is analogous
to ``noises'' in engineering according to the analysis of
\cite{bruit}.\footnote{The notion of ``noise'' has sometimes a quite
different meaning in quantitative finance (\cite{black}).} See
\cite{nl} and \cite{mboup} for the estimation of $S_{\tiny{\rm
trend}}(t)$ and of its derivatives. See \cite{agadir}, and the
references therein, for convincing numerical experiments including
forecasting results which are deduced from the trends.
\end{remark}
\subsection{Return}
If $S_{\tiny{{\mathfrak{TS}}, {\rm trend}}}$ is differentiable at
$t$, then its logarithmic derivative
\begin{equation}\label{return}
{\mathbf{r}}_{{\tiny{{\mathfrak{TS}}}}{, {\rm trend}}} (t) =
\frac{\dot{S}_{\tiny{{\mathfrak{TS}}, {\rm trend}}}
(t)}{S_{\tiny{{\mathfrak{TS}}, {\rm trend}}} (t)}
\end{equation}
is called the \emph{trend-return} of $S$ at $t$.
\begin{remark}
See \cite{douai,agadir} for other definitions of returns.
\end{remark}
\subsection{Volatility}
Take two integrable time series $S_1(t)$, $S_2 (t)$, such that their
squares and the squares of $S_{1, \tiny{\rm trend}}(t)$ and $S_{2,
\tiny{\rm trend}}(t)$ are also integrable. It leads us to the
following definitions, which are borrowed from \cite{douai,agadir}:
\begin{enumerate}
\item The \emph{covariance} of two time series $S_1(t)$ and $S_2(t)$ is
the time series
{\small
\begin{eqnarray*}
\mbox{\rm cov}(S_1 S_2)(t) & = & {\mathrm{Tr}}\left((S_1 - {\mathrm{Tr}}(S_1))(S_2 - {\mathrm{Tr}}(S_2)) \right)(t)
\\ & \simeq & {\mathrm{Tr}}(S_1 S_2)(t) - {\mathrm{Tr}}(S_1)(t) \times {\mathrm{Tr}}(S_2)(t)
\end{eqnarray*}
}
where ${\mathrm{Tr}} (\bullet)$ denotes the trend with respect to the
time sampling $\mathfrak{TS}$.
\item The \emph{variance} of the time series $S_1 (t)$ is
\begin{eqnarray*}\label{var}
\mbox{\rm var}(S_1)(t) & = & {\mathrm{Tr}}\left((S_1 - {\mathrm{Tr}}(S_1))^2 \right)(t) \\ & \simeq & {\mathrm{Tr}}(S_1^{2})(t) -
\left({\mathrm{Tr}}(S_1)(t)\right)^2
\end{eqnarray*}
\item The \emph{volatility} of $S_1(t)$ is the corresponding standard
deviation
\begin{equation}\label{vol}
\mbox{\rm vol}(S_1)(t) = \sqrt{\mbox{\rm var}(S_1)(t)}
\end{equation}
\end{enumerate}

\section{Pricing without trends}\label{arbi}
We limit ourselves for simplicity's sake to \emph{European call
options}, which are options for the right to buy a stock or an index
at a certain price at a certain maturity date.
\subsection{Arbitrage} Let $r (t)$ be the risk-free rate. The
expected price at maturity $T$ should be equal to
\begin{equation}\label{expec}
S  (0) \exp \left( \int_{0}^{T} r(\tau) d \tau \right)
\end{equation}
A heuristic justification goes like this: Assume, for simplicity's
sake and like in today's academic literature, that
\begin{itemize}
\item $r (t)$ is a constant $r$,
\item $S(t)$ follows a geometric Brownian motion
\begin{equation}\label{geometric}
S(t) = S(0) \exp \left[ \left(\mu - \frac{\sigma^2}{2} \right) t +
\sigma W(t) \right]
\end{equation}
where
\begin{itemize}
\item $W(t)$ is a standard Brownian motion,
\item $\mu$ and $\sigma$ are constant.
\end{itemize}
\end{itemize}
Providing a theoretical estimation of $\mu$ and $\sigma$ from
historical data is classic and straightforward. We thus know the
mean $S(0) e^{\mu t}$ of $S(t)$. If $\mu > r$ (resp. $\mu < r$), it
might be profitable for the arbitrageur to borrow money (resp.
selling the underlying) for buying the underlying $S$
(resp. for investing the corresponding amount of money) at time $0$,
and selling it (resp. buying the underlying) later, at time
$T$ for instance.
\subsection{Formulas} Assume that
\begin{itemize}
\item the underlying follows the geometric Brownian motion
\eqref{geometric},
\item the expected final price satisfies the condition \eqref{expec}, \textit{i.e.},
is equal to
$$S  (0) e^{rT}$$
\end{itemize}
Krouglov \cite{krouglov} shows, by exploiting properties of log-normal
distributions, that the usual BSM formulas may be recovered. Write
down here the value of a European call option:
\begin{equation}\label{BS1}
C(S, t) = S(t) N(d_1) - K N(d_2) e^{-r(T - t)}
\end{equation}
where
\begin{itemize}
\item $N$ is the standard normal cumulative distribution function,
\textit{i.e.},
$$
N(x) = \frac{1}{\sqrt{2\pi}} \int_{- \infty}^{x} \exp \left(-
\frac{z^2}{2} \right) dz
$$
\item $K$ is the strike price,
\item $d_1 = \frac{\lg\left(\frac{S}{K}\right) + \left(r + \frac{\sigma^2}{2} \right) (T - t)}
{\sigma \sqrt{T-t}}$,
\item $d_2 = d_1 - \sigma \sqrt{T - t}$.
\end{itemize}

\section{Pricing with trends}\label{wt}
\subsection{Arbitrage} Assume again that the risk-free rate $r(t)$
is a constant $r$. A natural extension of Section \ref{arbi} states
that the expected final price at maturity $T$ of the underlying
is
$$
S_{\tiny {\mathfrak{TS}}, {\rm trend}}  (0) e^{rT}
$$
It means the following:
\begin{itemize}
\item $S_{\tiny {\mathfrak{TS}}, {\rm trend}}  (0)$ replaces $S(0)$
in order to avoid the quick fluctuations.
\item The trend $S_{\tiny {\mathfrak{TS}}, {\rm trend}}  (t)$ is ``close''
around maturity $T$
to $S_{\tiny{\mathfrak{TS}}, {\rm trend}} (0) e^{rt}$.
\item The trend $S_{\tiny {\mathfrak{TS}}, {\rm trend}}  (t)$ is
differentiable around $T$ and the corresponding trend-return
${\mathbf{r}}_{{\tiny{{\mathfrak{TS}}}}{, {\rm trend}}} (t)$ of
Equation \eqref{return} is ``close'' to $r$.
\end{itemize}

\subsection{Formulas}\label{tf} Assume that the quick fluctuations around the
trend may be described at a time $t$ around $T$ by a lognormal
distribution of mean $S_{\tiny {\mathfrak{TS}}, {\rm trend}}  (t)$
and variance $\sigma$. It yields, as in Section \ref{arbi}, the
BSM-like formulas where the value of a European call option is given
by
\begin{equation}\label{BS2}
\boxed{C(S, t) = S_{\tiny {\mathfrak{TS}}, {\rm trend}} (t) N(d_1) -
K N(d_2) e^{-r(T - t)}}
\end{equation}
When compared to Equation \eqref{BS1}, notice that $S(t)$ is
replaced by $S_{\tiny {\mathfrak{TS}}, {\rm trend}} (t)$.

\begin{remark}
If we suppose that the quick fluctuations may be properly described
by a normal distribution, we would arrive at pricing formulas quite
analogous to those of \cite{bachelier} and
\cite{bronzin}.\footnote{Mimicking the computations with the other
probability distributions, which were considered by \cite{bronzin},
would be straightforward.} If we assume that we only forecast the
volatility \eqref{vol}, then the choice of the corresponding normal
distribution might be quite appropriate.
\end{remark}

\section{Dynamic Hedging}\label{dhed}

\subsection{General principles\protect\footnote{See \cite{delta} for
a related attempt.}} Let $\Pi$ be the value of an elementary
portfolio of one long option position $V$ and one short position in
quantity $\Delta$ of some underlying $S$:
\begin{equation}\label{Pi}
\Pi (t) = V (t) - \Delta S (t)
\end{equation}
Note that $\Delta$ is the control variable: the underlying is
sold or bought. The portfolio is {\em riskless} if its value obeys
the equation $ d \Pi = r \Pi dt $, where $r$ is the constant
risk-free rate. It yields
\begin{equation}\label{val}
\Pi (t) = \Pi (0) e^{rt}
\end{equation}
Replace
\begin{itemize}
\item Equation \eqref{Pi} by
\begin{equation}\label{Pit}
\Pi_{\tiny {\mathfrak{TS}}, {\rm trend}} (t) = V (t) - \Delta
S_{\tiny {\mathfrak{TS}}, {\rm trend}} (t)
\end{equation}
where $V$ is computed at time $t$ via Section \ref{tf}.
\item Equation \eqref{val} by
\begin{equation}\label{valt}
\Pi_{\tiny {\mathfrak{TS}}, {\rm trend}} (t) = \Pi_{\tiny
{\mathfrak{TS}}, {\rm trend}} (0) e^{rt}
\end{equation}
\end{itemize}
Combining Equations \eqref{Pit} and \eqref{valt} leads to the
tracking control strategy
\begin{equation}\label{hedging}
\boxed{\Delta = \frac{V (t) - \Pi_{\tiny {\mathfrak{TS}}, {\rm
trend}} (0) e^{rt}}{S_{\tiny {\mathfrak{TS}}, {\rm trend}} (t) }}
\end{equation}
We might again call {\em delta hedging} this strategy, although it
is only an approximate dynamic hedging via the utilization of trends
and of the corresponding time sampling ${\mathfrak{TS}}$.

In order to implement correctly Equation \eqref{hedging}, the
initial value $\Delta (0)$ of $\Delta$ has to be known. If $S_{\tiny
{\mathfrak{TS}}, {\rm trend}}$ and $V$ are differentiable, this is
achieved by equating the logarithmic derivatives at $t = 0$ of the
right handsides of Equations \eqref{Pit} and \eqref{valt}:
\begin{equation}\label{0}
\boxed{\Delta (0) = \frac{\dot{V} (0) - r V (0)}{\dot{S}_{\tiny
{\mathfrak{TS}}, {\rm trend}} (0) - r S_{\tiny {\mathfrak{TS}}, {\rm
trend}}  (0)}}
\end{equation}

\begin{remark}
Our approach to dynamic hedging may be connected to \emph{model-free control}
(\cite{esta,marseille}) which already found many concrete applications.\footnote{See the
references in \cite{marseille}.} Remember that one of the main difficulty related
to dynamic replication is the necessity to have a ``good'' probabilistic model of
the behavior of the underlying.
\end{remark}

\section{Some computer illustrations}\label{illu}
The underlying is the S\&P 500, which is one of the most commonly
followed equity indices.
\subsection{Preliminary calculations}\label{prelim}
The preliminary calculations below are necessary for our dynamic
hedging in Section \ref{dh}.

\subsubsection{Data and trends} 
\begin{figure}
\begin{center}{\rotatebox{-0}{\includegraphics*[width=1.1\columnwidth]{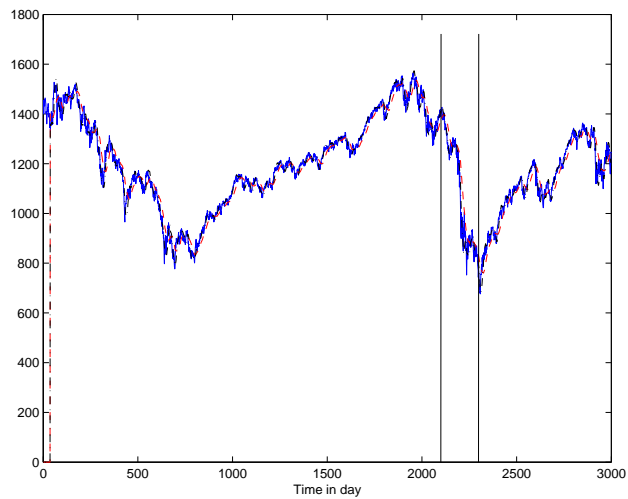}}}\end{center}
 \caption{S\&P 500 value (blue, --), its moving average (red, - -) and the proposed trend (black, .-) \label{F1}}
\end{figure}

\begin{figure}
\begin{center}{\rotatebox{-0}{\includegraphics*[width=1.1\columnwidth]{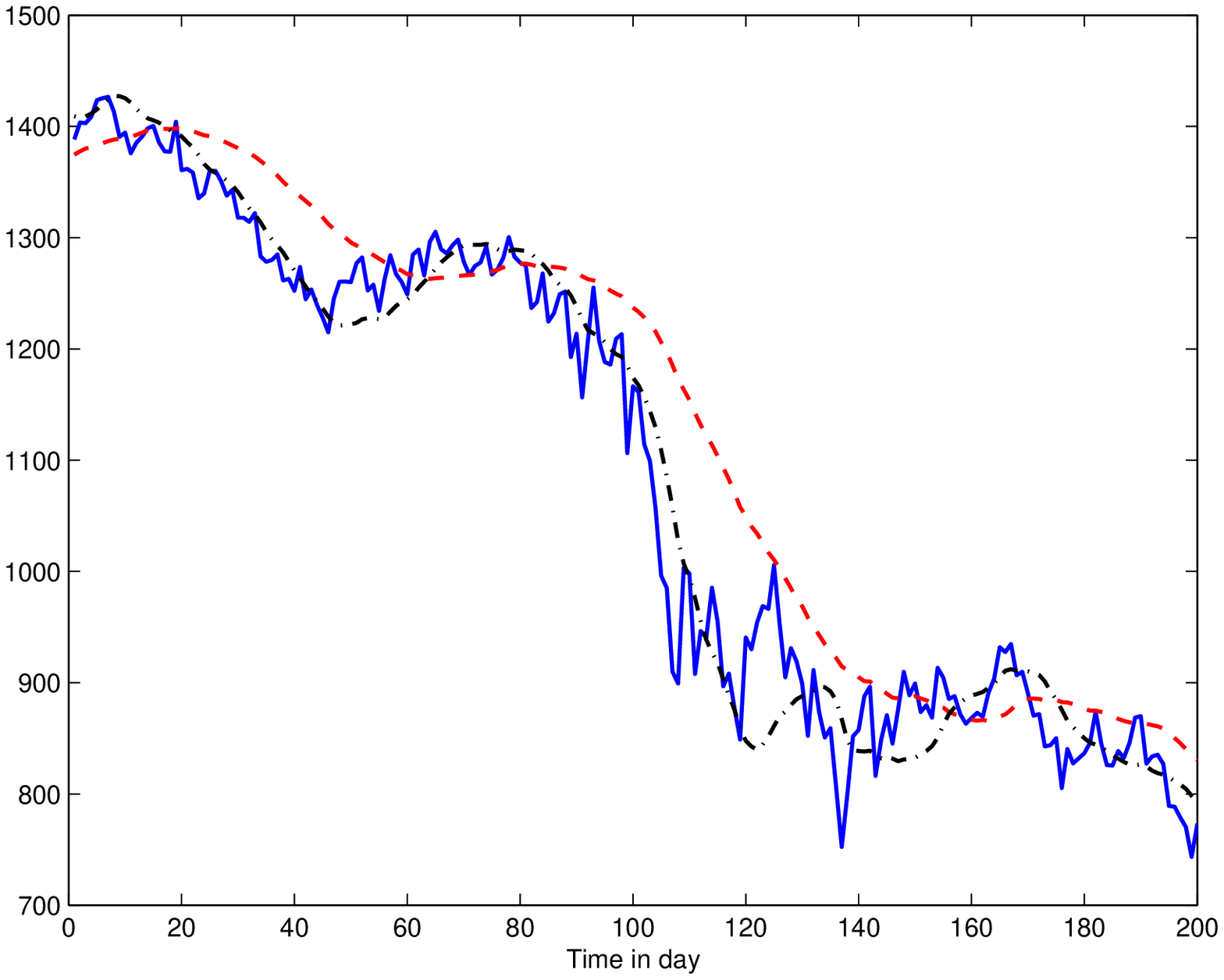}}}\end{center}
 \caption{Zoom of Figure \ref{F1} \label{F2}}
\end{figure}

Figure \ref{F1} displays the daily
S\&P 500, from 3 January 2000 until 2 December 2012. A turbulent 200
days period  from 9 May 2008 until 24 February 2009 is extracted in
Figure \ref{F2}. The excellent quality of our trend estimation (see
Remark \ref{trendagadir}) is highlighted by those two Figures,
especially when compared to a classic moving average techniques
using the same number of points, here 30. Let us emphasize moreover
that the unavoidable delay associated to any estimation technique is
quite reduced thanks to our theoretical viewpoint.

\subsubsection{Volatility}\label{vola} 
\begin{figure}
\begin{center}{\rotatebox{-0}{\includegraphics*[width=1.1\columnwidth]{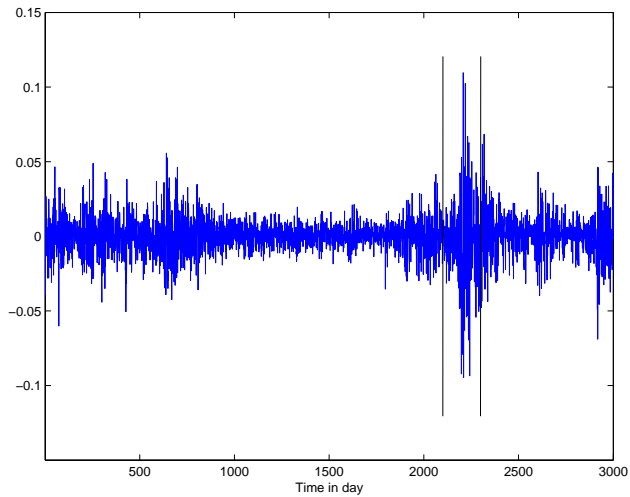}}}\end{center}
 \caption{Log return \label{F3}}
\end{figure}

\begin{figure}
\begin{center}{\rotatebox{-0}{\includegraphics*[width=1.1\columnwidth]{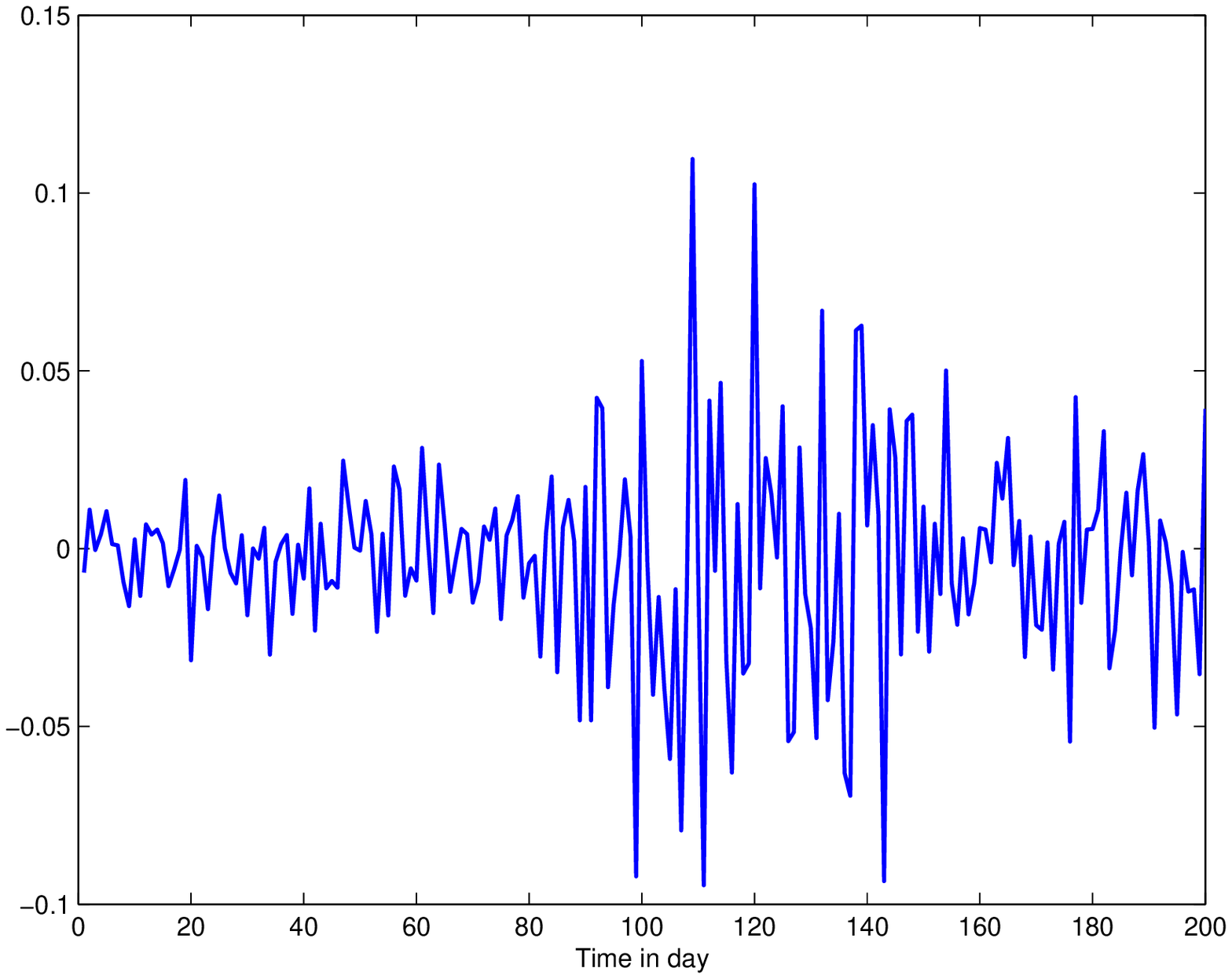}}}\end{center}
 \caption{Zoom of Figure \ref{F3}  \label{F4}}
\end{figure}

\begin{figure}
\begin{center}{\rotatebox{-0}{\includegraphics*[width=1.1\columnwidth]{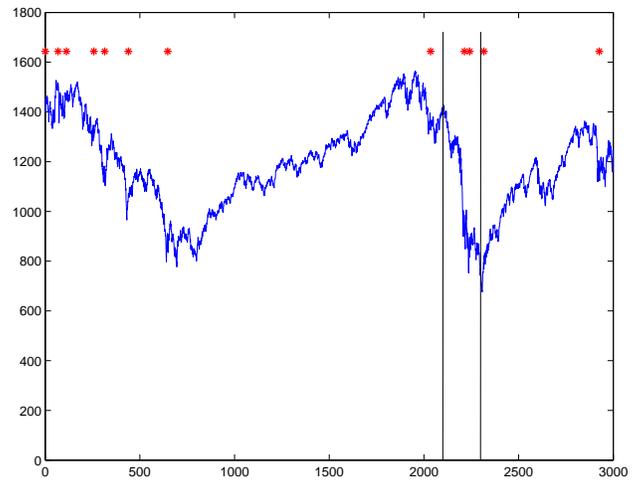}}}\end{center}
 \caption{Abrupt change detection (red,$*$) on S\&P 500  \label{F5}}
\end{figure}

\begin{figure}
\begin{center}{\rotatebox{-0}{\includegraphics*[width=1.1\columnwidth]{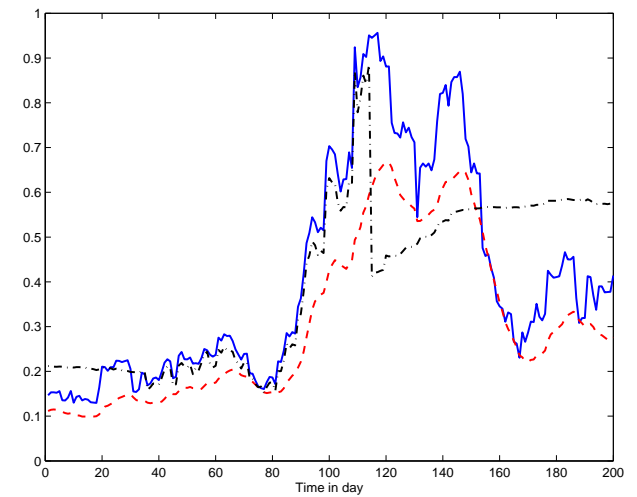}}}\end{center}
 \caption{Usual volatility (blue, --), filtered volatility (red, - -), and time-scaled volatility (black, . -)\label{F6}}
\end{figure}

\begin{figure}
\begin{center}{\rotatebox{-0}{\includegraphics*[width=1.1\columnwidth]{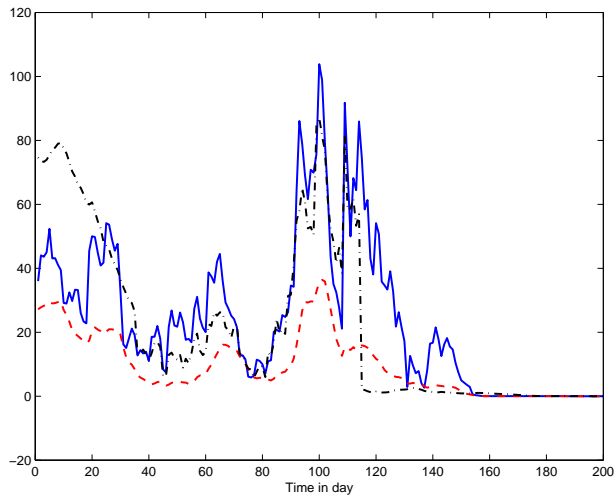}}}\end{center}
 \caption{Call (blue, --), call with filtered volatility (red, - -) and call with time-scaled volatility (black, . -) \label{F7}}
\end{figure}

Figure \ref{F3} and \ref{F4} display the
corresponding logarithmic return
$$R(t)=\ln \left(\frac{S(t)}{S(t-1)} \right)$$
where $S(t)$ denotes the daily value of the S\&P 500 and $t>1$. The
corresponding annualized volatility is
$$\sigma(t)= \text{STD}(R(t)) \times \sqrt{255}$$
where, for determining the standard deviation $\text{STD}$,
\begin{itemize}
\item a 10 days sliding window is used,
\item the mean may be deduced from Equation \eqref{return}.
\end{itemize}
This type of calculations is much too sensitive to the return
fluctuations. Figure \ref{F6} exhibits this annoying feature as well
as the results obtained via the two following procedures which are
utilized in order to bypass this difficulty:
\begin{enumerate}
\item A classic low-pass filter permits to alleviate those fluctuations.

\item The results for the on-line detection methods in \cite{abrupt} of
\emph{change-points}\footnote{This terminology, which is borrowed
from the literature on signal processing (see \cite{abrupt} and the
references therein), seems more appropriate than the word
\emph{jumps} which is familiar in quantitative finance.} are
depicted in Figure \ref{F5}. The sensitivity of the algorithm, which
may be easily modified, is adapted here to quite violent abrupt
changes. If such a change is detected its effect is reduced via
an averaging where the size of the sliding window is augmented. It corresponds to the
\emph{time-scaled volatility} in Figures \ref{F6}, \ref{F7} and \ref{F8}.
\end{enumerate}

The second method, which provides a most efficient smoothing when a
change point is detected, seems to work better.

\subsubsection{Option pricing} Introduce now the European call option
during the hectic period of 200 days shown in Figure \ref{F2}. Write $T
= 200$ the maturity time. Set $r=1\%$ for the risk-free rate.
The strike price $K$ is given by
$$K=S_{\tiny{\mathfrak{TS}}, {\rm
trend}}(0)(k/100+1)^{(T/255)}$$
where $k=10\%$. At any time $t$, $0
< t < T$, computing the numerical value of the call, as shown in
Figure \ref{F7}, uses
\begin{itemize}
\item Formula \eqref{BS2},\footnote{Only lack of space
makes us follow here a Black-Scholes type formula.}
\item the estimated volatilities in Section \ref{vola}.
\end{itemize}

\subsection{Dynamic hedging}\label{dh}
\begin{figure}
\begin{center}{\rotatebox{-0}{\includegraphics*[width=1.1\columnwidth]{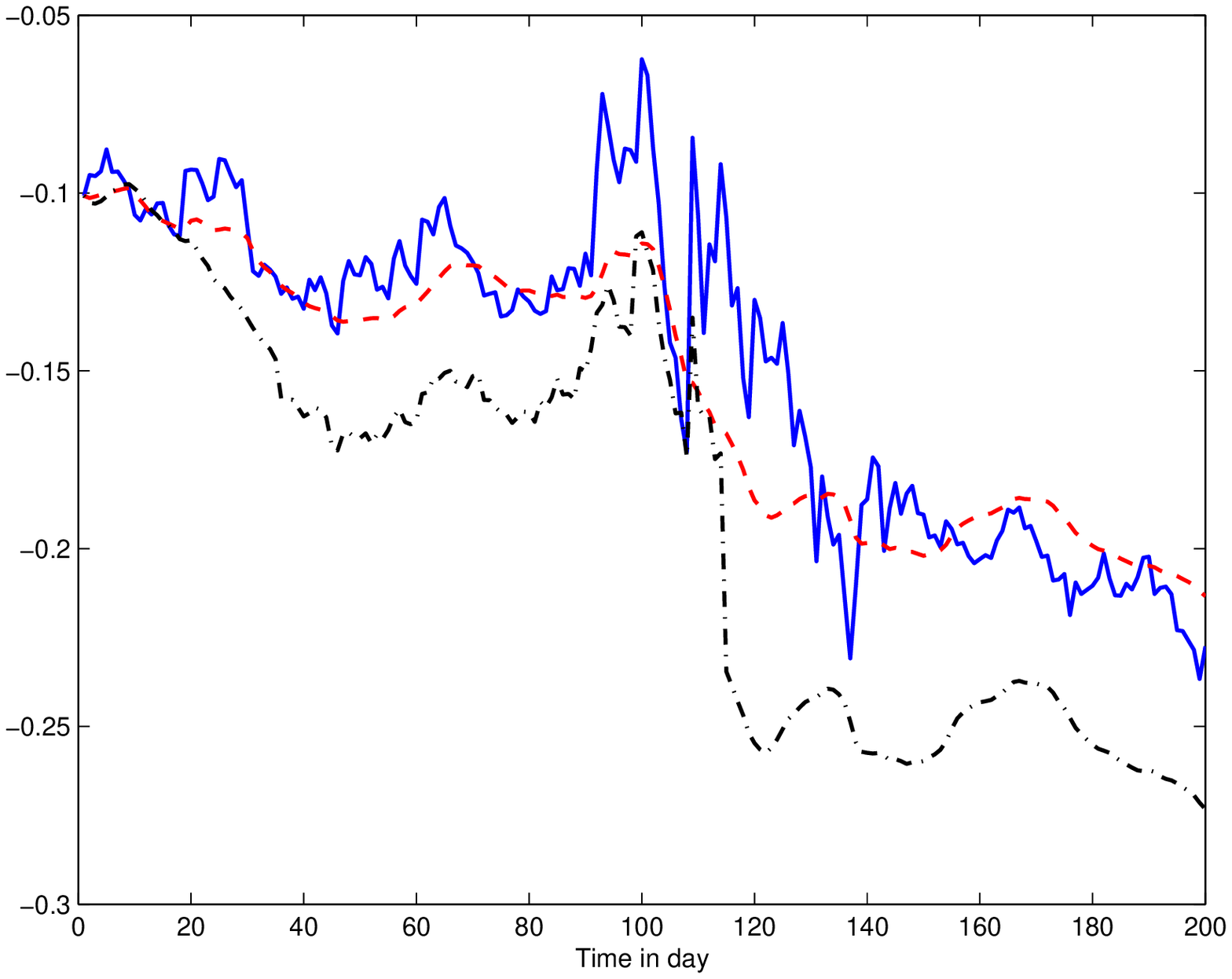}}}\end{center}
 \caption{Hedging (blue, --), hedging with filtered volatility (red, - -), and hedging with time-scaled volatility (black, . -)  \label{F8}}
\end{figure}

Thanks to the numerical results of Section \ref{prelim},
Formula \eqref{hedging} yields dynamic hedging performances which are reported
in Figure \ref{F8}. Note that a proper choice of the volatility calculation
ensures in the same time and in spite of an only rough replication
\begin{itemize}
\item small oscillations of the control variable $\Delta$,
\item a good hedging.
\end{itemize}



\section{Conclusion}\label{conclu}
If further studies confirm our viewpoint on option pricing and dynamic hedging,
it will open radically different roads which should bypass some of the most important difficulties
encountered with today's approaches. Let us emphasize as above and once
again (\cite{fes,agadir}) that
a consequence of our setting might the obsolescence
of the need of complex stochastic processes for modeling the underlying's behavior.
Taking into account
\begin{itemize}
\item the trends, which carry the information about jumps and other ``violent'' behaviors,
\item their forecasting,
\item not only the variance around the trend but also the skewness and the kurtosis,
\end{itemize}
should lead to new option pricing formulas, where the (geometric) Brownian motion
will loose its preeminence.

\emph{American} and other \emph{exotic} options will be considered elsewhere.

\section*{Acknowledgment}
The authors thank Frank G\'{e}not and Fr\'{e}d\'{e}ric Hatt for helpful
discussions.



%

\end{document}